\documentclass[12pt,preprint]{aastex}

\def\Prob   {\ifmmode {{\rm Prob}} \else {Prob} \fi}
\def\arm    {\ifmmode {{\rm arm}} \else {arm} \fi}

\def\twopi  {\ifmmode {2\pi}   \else {$2\pi$}   \fi}

\def\kms    {\ifmmode{{\rm km s}^{-1}}\else{km s$^{-1}$}\fi}

\def\uas  {\ifmmode {\mu{\rm as}}\else{$\mu$as}\fi}
\def\deg  {\ifmmode {^\circ}\else {$^\circ$}\fi}
\def\porm {\ifmmode {\pm}\else {$\pm$}\fi}

\def\chisqpdf {\ifmmode {\chi^2_{\rm pdf}}\else {$\chi^2_{\rm pdf}$}\fi}
\def\chisq    {\ifmmode {\chi^2}\else {$\chi^2$}\fi}

\def\etal {et al.~}
\def\eg   {e.g.,~}
\def\ie   {i.e.,~}

\def\d    {\ifmmode {{\rlap{.}}^\circ}\else {${\rlap{.}}^\circ$}\fi}
\def\s    {\ifmmode {{\rlap{.}}^s}\else {${\rlap{.}}^s$}\fi}
\def\as   {\ifmmode {{\rlap{.}}^{''}}\else {${\rlap{.}}^{''}$}\fi}

\newbox\grsign \setbox\grsign=\hbox{$>$} \newdimen\grdimen \grdimen=\ht\grsign
\newbox\laxbox \newbox\gaxbox
\setbox\gaxbox=\hbox{\raise.5ex\hbox{$>$}\llap
     {\lower.5ex\hbox{$\sim$}}}\ht1=\grdimen\dp1=0pt
\setbox\laxbox=\hbox{\raise.5ex\hbox{$<$}\llap
     {\lower.5ex\hbox{$\sim$}}}\ht2=\grdimen\dp2=0pt
\def\gax{\mathrel{\copy\gaxbox}}
\def\lax{\mathrel{\copy\laxbox}}

\def\pa    {\ifmmode {\psi} \else {$\psi$}\fi}
\def\rPpm  {\ifmmode {r_{\Ro,\To}} \else {$r_{Ro,\To}$}\fi}

\def\vlsr  {\ifmmode {v}\else {$v$}\fi}
\def\vhelio{\ifmmode {v_{Helio}}\else {$v_{Helio}$}\fi}
\def\delV  {\ifmmode {\Delta v}\else {$\Delta v$}\fi}
\def\sigV  {\ifmmode {\sigma_v}\else {$\sigma_v$}\fi}

\def\ura   {\ifmmode {\mu_\alpha}\else {$\mu_\alpha$}\fi}
\def\udec  {\ifmmode {\mu_\delta}\else {$\mu_\delta$}\fi}

\def\ul    {\ifmmode {\mu_l}\else {$\mu_l$}\fi}
\def\ub    {\ifmmode {\mu_b}\else {$\mu_b$}\fi}

\def\uml   {\ifmmode {v_{gr}}\else {$v_{gr}$}\fi}
\def\umb   {\ifmmode {v_b}\else {$v_b$}\fi}
\def\vsrad {\ifmmode {v_{rad}}\else {$v_{rad}$}\fi}

\def\upl   {\ifmmode {v^p_{gr}}\else {$v^p_{gr}$}\fi}
\def\upb   {\ifmmode {v^p_b}\else {$v^p_b$}\fi}
\def\vprad {\ifmmode {v^p_{rad}}\else {$v^p_{rad}$}\fi}

\def\Vo    {\ifmmode {V^{Std}_\odot}\else {$V^{Std}_\odot$}\fi}
\def\Uo    {\ifmmode {U^{Std}_\odot}\else {$U^{Std}_\odot$}\fi}
\def\Wo    {\ifmmode {W^{Std}_\odot}\else {$W^{Std}_\odot$}\fi}
\def\VH    {\ifmmode {V^H_\odot}\else {$V^H_\odot$}\fi}
\def\UH    {\ifmmode {U^H_\odot}\else {$U^H_\odot$}\fi}
\def\WH    {\ifmmode {W^H_\odot}\else {$W^H_\odot$}\fi}
\def\V     {\ifmmode {V_\odot}\else {$V_\odot$}\fi}
\def\U     {\ifmmode {U_\odot}\else {$U_\odot$}\fi}
\def\W     {\ifmmode {W_\odot}\else {$W_\odot$}\fi}

\def\Vs    {\ifmmode {V_s}\else {$V_s$}\fi}
\def\Us    {\ifmmode {U_s}\else {$U_s$}\fi}
\def\Ws    {\ifmmode {W_s}\else {$W_s$}\fi}

\def\Vsbar {\ifmmode {\overline{V_s}}\else {$\overline{V_s}$}\fi}
\def\Usbar {\ifmmode {\overline{U_s}}\else {$\overline{U_s}$}\fi}
\def\Wsbar {\ifmmode {\overline{W_s}}\else {$\overline{W_s}$}\fi}

\def\aone  {\ifmmode {a_1}\else {$a_1$}\fi}
\def\atwo  {\ifmmode {a_2}\else {$a_2$}\fi}
\def\athr  {\ifmmode {a_3}\else {$a_3$}\fi}

\def\pars  {\ifmmode{\pi_s}\else{$\pi_s$}\fi}

\def\Ts    {\ifmmode{\Theta_s}\else{$\Theta_s$}\fi}

\def\To    {\ifmmode{\Theta_0}\else{$\Theta_0$}\fi}
\def\T     {\ifmmode{\Theta}\else{$\Theta$}\fi}
\def\Ro    {\ifmmode{R_0}\else{$R_0$}\fi}

\def\Tdot  {\ifmmode{\dot{\Theta}}\else{$\dot{\Theta}$}\fi}
\def\Tddot {\ifmmode{\ddot{\Theta}}\else{$\ddot{\Theta}$}\fi}

\def\lbv     {\ifmmode {(l,b,v)}\else{$(l,b,v)$}\fi}
\def\lv      {\ifmmode {(l,v)}\else{$(l,v)$}\fi}
\def\lvS     {\ifmmode {(l,v)_{\rm src}}\else{$(l,v)_{\rm src}$}\fi}
\def\lbvS    {\ifmmode {(l,b,v)_{\rm src}}\else{$(l,b,v)_{\rm src}$}\fi}
\def\lbvA    {\ifmmode {(l,b,v)_{\rm arm}}\else{$(l,b,v)_{\rm arm}$}\fi}
\def\lbvRBD  {\ifmmode {(l,b,v,R,\beta,d)}\else{$(l,b,v,R,\beta,d)$}\fi}

\def\Nbins   {\ifmmode{N_{\rm bins}}\else{$N_{\rm bins}$}\fi}
\def\DelD    {\ifmmode{\Delta d}\else{$\Delta d$}\fi}

\def\Rtp     {\ifmmode{R_{tp}}\else{$R_{tp}$}\fi}
\def\Vtp     {\ifmmode{V_{tp}}\else{$V_{tp}$}\fi}
\def\Vx      {\ifmmode{V_x}\else{$V_x$}\fi}

\def\Qij     {\ifmmode{(x^j_i,y^j_i)}\else{$(x^j_i,y^j_i)$}\fi}
\def\Qxij    {\ifmmode{x^j_i}\else{$x^j_i$}\fi}
\def\Qyij    {\ifmmode{y^j_i}\else{$y^j_i$}\fi}

\def\Offset  {\ifmmode{(\Theta_x^j,\Theta_y^j)}\else{$(\Theta_x^j,\Theta_y^j)$}\fi}
\def\Offsetx {\ifmmode{\Theta_x^j}\else{$\Theta_x^j$}\fi}
\def\Offsety {\ifmmode{\Theta_y^j}\else{$\Theta_y^j$}\fi}

\def\grad    {\ifmmode{{\partial p}\over{\partial\theta}}\else{${\partial p}\over{\partial\theta}$}\fi}

\slugcomment{2017 June 5}
\shorttitle{Parallax Measurement Techniques} 
\shortauthors{Reid et al.}

\begin{document}

\title{Techniques for Accurate Parallax Measurements for 6.7-GHz
       Methanol Masers}   

\author{M. J. Reid\altaffilmark{1}, 
        A. Brunthaler\altaffilmark{2}, K. M. Menten\altaffilmark{2},
        A. Sanna\altaffilmark{2},
        Y. Xu\altaffilmark{3}, J. J. Li\altaffilmark{3},
        Y. Wu\altaffilmark{3}, B. Hu\altaffilmark{3},
        X. W. Zheng\altaffilmark{4},
        B. Zhang\altaffilmark{5},
        K. Immer\altaffilmark{6},
        K. Rygl\altaffilmark{7},
        L. Moscadelli\altaffilmark{8},
        N. Sakai\altaffilmark{9},
        A. Bartkiewicz\altaffilmark{10},
        Y. K. Choi\altaffilmark{11}
       }

\altaffiltext{1}{Harvard-Smithsonian Center for
   Astrophysics, 60 Garden Street, Cambridge, MA 02138, USA}
\altaffiltext{2}{Max-Planck-Institut f\"ur Radioastronomy,
   Auf dem H\"ugel 69, 53121-Bonn, Germany}
\altaffiltext{3}{Purple Mountain Observatory, Chinese Academy of Sciences, 
   Nanjing 210008, China}
\altaffiltext{4}{Department of Astronomy, Nanjing University, Nanjing 210093, 
   China}
\altaffiltext{5}{Shanghai Astrophysical Observatory, 80 Nandan Rd, Shanghai 200030, 
   China}
\altaffiltext{6}{European Southern Observatory, Karl-Schwarzschild-Strasse 2, 
   85748, Garching bei M\"unchen, Germany}
\altaffiltext{7}{Italian ALMA Regional Centre, INAF -- Istituto di Radioastronomia, 
   Via P. Gobetti 101, 40129 Bologna, Italy}
\altaffiltext{8}{INAF-Osservatorio Astrofisico di Arcetri, Largo E. Fermi 5, 
   50125, Firenze, Italy}
\altaffiltext{9}{Mizusawa VLBI Observatory, National Astronomical Observatory 
   of Japan, 2-21-1 Osawa, Mitaka, Tokyo 181-8588, Japan}
\altaffiltext{10}{Centre for Astronomy, Faculty of Physics, Astronomy and 
   Informatics, Nicolaus Copernicus University, Grudziadzka 5, 87-100 Torun, Poland}
\altaffiltext{11}{Korea Astronomy and Space Science Institute 776, Daedeokdae-ro, 
   Yuseong-gu, Daejeon, Republic of Korea (34055)}

\begin{abstract}
The BeSSeL Survey is mapping the spiral structure of the Milky Way
by measuring trigonometric parallaxes of hundreds of maser sources
associated with high-mass star formation.   While parallax techniques for
water masers at high frequency (22 GHz) have been well documented, recent 
observations of methanol masers at lower frequency (6.7 GHz) have revealed 
astrometric issues associated with signal propagation through the ionosphere 
that could significantly limit parallax accuracy.  These problems displayed
as a ``parallax gradient'' on the sky when measured against different background
quasars.  We present an analysis method in which we generate position data
relative to an ``artificial quasar'' at the target maser position at each epoch.
Fitting parallax to these data can significantly mitigate the problems and 
improve parallax accuracy.
\end{abstract}


\section{Introduction} \label{sect:introduction}

The Bar and Spiral Structure Legacy (BeSSeL) Survey\footnote[12]{\tt http://bessel.vlbi-astrometry.org}
uses the National Radio Astronomy Observatory's\footnote[13]{The National Radio Astronomy Observatory is a facility of the National Science Foundation operated under cooperative agreement by Associated Universities, Inc.}  
Very Long Baseline Array (VLBA) to 
measure parallaxes of methanol and water masers associated with newly 
formed (or forming) high-mass stars throughout the Milky Way.  In the first 
two years of the survey, observations were done using maser lines near 12 GHz 
(methanol) and 22 GHz (water).  At frequencies above $\approx10$ GHz, 
the dominant source of astrometric error is usually uncompensated interferometric 
delays associated with water vapor in the troposphere.  Astrometric techniques 
that can yield $\sim10$ micro-arcsec parallaxes at these frequencies are 
described in \citet{Honma:08}, \citet{Reid:09a} and \citet{Reid:14b}.

However, only about a dozen 12-GHz methanol masers were strong enough 
to serve as interferometer phase-reference sources and hence be optimum 
targets for parallax measurements, and the BeSSeL Survey sought to use the
stronger and more numerous 6.7-GHz methanol masers.  
Since emission line features of methanol masers are longer-lived 
(typically decades) than water masers (often only several months), 
interferometer coherence times improve with observing wavelength, and 
atmospheric opacity is generally lower at longer cm-wavelengths, we 
anticipated that parallax accuracy would be comparable to or better than those 
obtained at 22 GHz.  So, in 2015, new wide C-band receivers, funded by 
the Max Planck Institute for Radio Astronomy (MPIfR), were installed on 
the VLBA.  Unfortunately, below $\approx10$ GHz, interferometric propagation 
delays associated with electrons in the ionosphere can severely increase
astrometric errors, even after removing estimated ionospheric delays
based on global total electron content models derived from Global Positioning
System data \citep{Walker:99}.  

The BeSSeL Survey strategy has been to use the target maser as
the interferometric phase-reference source and rapidly switch
between the maser and background quasars, in order to remove short-term
phase-delay fluctuations (mostly from water vapor) from the interferometric 
data.  This essentially removes interferometer coherence-time limitations 
and allows one to use Earth rotation synthesis to improve interferometer 
(u,v)-coverage, imaging quality, and astrometric accuracy.  We used 
several background quasars for each maser, to guard against occasional 
structural variability in a quasar limiting astrometric accuracy, as well 
as to provide several nearly independent parallax measurements.  

Often the quasars surrounded the target maser on the sky.   This was
fortunate as it allowed us to detect systematic gradients in parallax
estimates on the sky as sampled by the different quasars relative to
the target maser source.  Presumably these can be traced to uncompensated 
ionospheric delays that present as ``wedges,'' which over time-scales of 
hours distort relative position measurements in a quasi-linear fashion 
over $\gax5^\circ$ on the sky.   At any single epoch it would be difficult 
to disentangle these effects from catalog position errors, which are 
generally $\gax0.1$ mas.   However, for parallax measurements which involve
multiple epochs spanning one year, this proved obvious as will be
shown below.   Since we do not see these effects at 22 GHz, which
is a high enough frequency that residual dispersive ionospheric delays 
should be small (\eg residual path-delays $\lax1$ cm), it is nearly 
certain that the astrometric problems seen at 6.7 GHz can be traced to
the propagation of the maser signal through the ionosphere.
 
This paper describes the challenges faced when observing at 6.7 GHz 
and the techniques we used to minimize astrometric errors.
In Section \ref{sect:observed} we describe the astrometric problems
evident in our data.   In Section \ref{sect:methods} we present a
calibration technique that significantly improves parallax accuracy.
Finally, in Section \ref{sect:considerations} we discuss the 
implications of the technique for the estimation of parallax uncertainty. 
 
\section{Observed Astrometric Problems} \label{sect:observed}

The equipment setup and calibration procedures for previous BeSSeL Survey 
observations at 12 and 22 GHz are documented in \citet{Reid:09a}.   
Importantly, for this paper, we use global models of the ionosphere's total 
electron content to remove an estimate of the dispersive delay along 
a ray-path toward any source.  The application of these models reduces residual 
dispersive delays by a factor of between two and five \citep{Walker:99}.
While this is generally adequate to reduce dispersive path-delays to 
$\sim1$ cm for 22-GHz observations, since dispersive delays scale as 
$\nu^{-2}$, where $\nu$ is the observing frequency, at 6.7 GHz residual 
path-delays can be $\sim10$ cm.  This is roughly an order of magnitude larger 
than our target accuracy of $\approx1$ cm, which is necessary to achieve
parallax accuracy of $\approx0.01$ mas \citep{Reid:14b}. 
If this not dealt with, it could limit parallax accuracy to $\sim0.1$ mas 
and, thus, limit distance measurement to $\lax1$ kpc with 10\% accuracy.  

\begin{figure}[ht]
\epsscale{1.0} 
\plotone{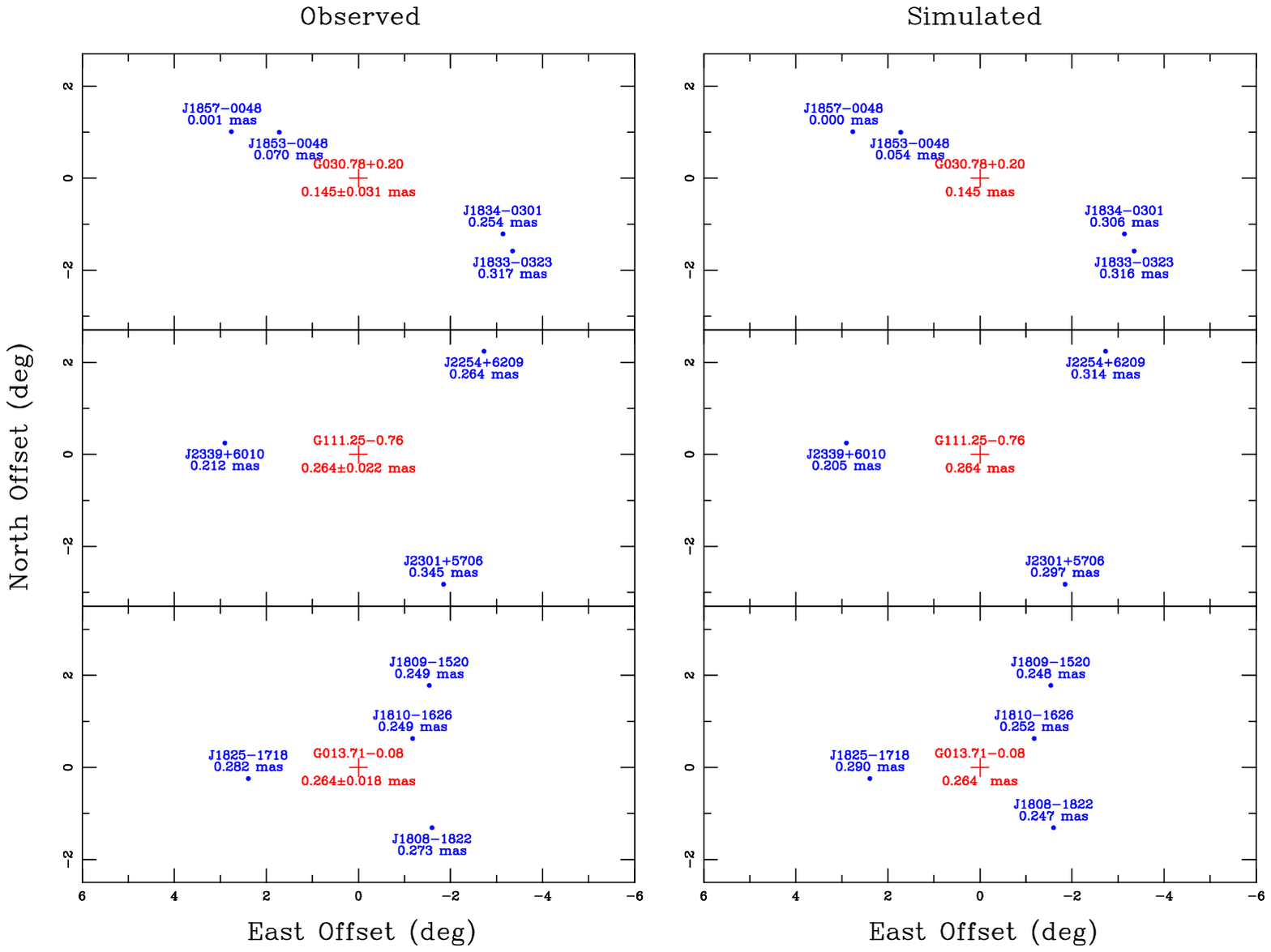}
\caption{\footnotesize
Locations of quasars (filled circles) which served as background 
reference sources for parallax measurements of three different Galactic 
6.7-GHz methanol masers (red plus signs).  
{\it Left Panels:} Individual parallax estimates
for the maser relative to each quasar are indicated below the quasar name.
The final parallax for the maser using Method-2 to generate an artificial
quasar at the position of the maser is given below the maser name.
{\it Right Panels:} Simulations of parallax measurement for the source groupings, 
assuming that relative position measurements
have been affected by systematic gradients across the sky.  The simulations 
assume relative position measurements are shifted in a planar fashion with
a gradient of 0.1 mas per degree of offset on the sky, with randomly different
orientations for each of four epochs used to fit parallax.  
{\it Top Panels:} Results for the maser G030.76+0.20.  The quasars
straddle the maser in a nearly linear arrangement and show a large parallax
gradient across their distribution.
{\it Middle Panel:} Results for the maser G111.25-0.76.  There is
a moderate gradient in the individual quasar parallax estimates from northeast to southwest.
{\it Bottom Panel:} Results for the maser G013.71-0.08.  This is
a case with little or no gradient in parallax.
        }
\label{fig:parallax_on_sky}
\end{figure}

The left panels of Fig.~\ref{fig:parallax_on_sky} show a range of examples for 
parallaxes of 6.7-GHz masers as measured against different quasars as a function of the 
locations of the quasars relative to the Galactic maser.  Of course, being at very 
great distances, all quasars used as background sources should yield the same parallax 
for the target maser within measurement uncertainty.  However, separate parallax 
measurements for a maser based on different quasars typically showed a ``parallax gradient'' 
on the sky with lower parallaxes for quasars on one side of the maser and larger 
parallaxes for quasars on the other side.  These parallax differences usually significantly 
exceeded expected parallax uncertainties.   Also, parallaxes measured against quasars 
that projected close together on the sky (\eg within $\approx1^\circ$) generally yielded 
similar values.   All of these findings argue against quasar structural variability as 
a significant contributing factor to parallax differences.  

We hypothesize that at each epoch uncompensated propagation delays resulted 
in systematic errors in the measured {\it relative} position shifts between 
maser and quasar, and, even if these errors are uncorrelated at different epochs, 
they affect the parallax and proper motion estimates, especially when only a 
small number of observations are used.  In order to test this hypothesis,
we simulated the effects of a planar relative-position ``wedge,'' characterized
by a position gradient on the sky, \grad, oriented at position angle East of North, 
$\phi$.  This results in a relative (East,North) position shift of 
$(\delta x,\delta y)=(\grad~\sin{\phi}~\Theta_x,\grad~\cos{\phi}~\Theta_y),$
where ($\Theta_x,\Theta_y$) is the separation of a background quasar from
the target maser source.

We assumed that the position angles of the wedges were uncorrelated 
among observing epochs and chose different values at each epoch from a uniform 
random distribution.  We then fitted these simulated position shifts for
our standard four-epoch observing sequence, which symmetrically sampled the peaks 
of the Right Ascension parallax signature (see Section \ref{sect:considerations}
for an example).  After some experimentation, we found that a characteristic value 
$\grad = 0.1$ mas deg$^{-1}$ 
yielded parallax gradients similar to those we observed.
Adopting this value, we changed the random number generator seed used to
set the position angles of the wedges and simulated parallax measurements for each 
of the three configurations of sources.  In the right panels of 
Fig. \ref{fig:parallax_on_sky}, we show examples that roughly match
the real observations.  Typically, we needed less than five simulations to find
such matches, with each simulation giving two results by reversing the signs
of the parallax offsets (\ie corresponding to a 180$^\circ$ rotation of the 
position angles of the wedges).  We conclude that a position gradient on the sky 
of $\sim0.1$ mas deg$^{-1}$ with a random orientation at each epoch can explain 
the variation in our 6.7-GHz parallax results of $\approx0.01$ to $\approx0.05$ 
mas deg$^{-1}$ for a maser source relative to background quasars.

As discussed earlier, there is strong circumstantial evidence that uncompensated 
ionospheric delays are at the root of the relative position wedges.  However, the 
details of how this happens are not clear.  The VLBA antennas span about 90$^\circ$ 
in longitude, or six hours of solar hour-angle, which should result in significant 
differences in ionospheric conditions above many of the antennas.  Thus, one might 
expect that the effects of ionospheric delays would be only partially correlated 
among the different sites and would somewhat ``average out'' in an image made
with the entire array over a 6-hour observation.  However, it is likely 
that the total electron content models used to remove most of the ionospheric 
delays may systematically over or under estimate the electrons, leading to 
significant correlations in the {\it residual} (uncompensated) delays and 
enhancing relative position shifts.  While this should be investigated in 
the future, it is beyond the scope of this paper, which is primarily concerned 
with a mitigation strategy to improve parallax accuracy regardless of the 
cause of the problem.

\section{Analysis Methods} \label{sect:methods}

In order to deal with the problem of position gradients across our sources, 
we take advantage of their distribution on the sky.   
\citet{Rioja:17} recently demonstrated a technique called ``MultiView'' 
that uses a two-dimensional interpolation of interferometer phase from a 
distribution of sources to calibrate phase towards a central 
target source.  They have shown impressive astrometric results at a frequency 
of 1.6 GHz, where ionospheric effects are more severe than for our observations
at 6.7 GHz.  Unfortunately, the BeSSeL Survey observations were not conducted 
in a manner that allows direct use of MultiView calibration.  The background 
quasars were selected to be close to the target, which generally required weak 
sources that could only be detected in images that combined data from multiple 
baselines and spanning hours of observation.  However, we can use a 
variant of MultiView with positions measured from images to improve our 
astrometric accuracy.

Conceptually, one could take the relative position measurements (maser minus 
quasar) at each epoch and estimate what would be the measurement relative to an 
``artificial quasar'' near or at the position of the maser, and then fit the 
artificial quasar positional data to estimate parallax and proper motion of
the maser.   The simplest approach would be to average the positions of all 
quasars (relative to the maser) to generate the artificial quasar data at each 
epoch.  However, this would not take into account the distribution of the quasars 
with respect to the maser.  Given the quasi-linear behavior on the sky of the 
parallax results, a better method is to allow for a gradient in relative position 
as a function of separation from the maser.   We investigated two such methods.  

Method-1 fits a plane through the 2-dimensional relative-position 
measurements\footnote[14]{To avoid potential confusion, we use the term 
``separations'' to refer to the degree-scale separations of individual quasars 
from a maser target, and the term ``positions'' to refer to the measured 
mas-scale position differences of the quasars from the maser after removing the 
degree-scale separations.}.  
Specifically, given the measured position of the $j^{th}$ quasar 
relative to the maser at epoch $i$, \Qij, and the separation on the sky of that 
quasar relative to the maser, \Offset, we fitted the components of the measured 
positions with the model
$$\Qxij = S^x_x\Offsetx + S^x_y\Offsety + C^x~~,\eqno(1)$$
and
$$\Qyij = S^y_x\Offsetx + S^y_y\Offsety + C^y~~.\eqno(2)$$
In Eqs.~~(1) and (2), the $S$ (slope) parameters allow for a tilt of the
plane and the $C$ (constant) parameters give the estimated position of the
artificial quasar at the maser position.  The superscript for the $S$ parameters
indicates the component of data being modeled, whereas the subscript indicates
the direction of separations for which the slope parameter applies.  Fitting this 
model requires at least three quasars, since for each direction on the sky there 
are three parameters.  A drawback of this method is that the parameters 
can become degenerate as the quasars approach a linear distribution on the sky.

\begin{deluxetable}{lrlrrr}
\tablecolumns{4} \tablewidth{0pc} 
\tablecaption{Parallax Results by Method}
\tablehead {
 \colhead{Source} &\colhead{Standard Parallax} &\colhead{Method-2 Parallax} \\
 \colhead{}       &\colhead{(mas)}             &\colhead{(mas)}      
           }
\startdata
G030.78$+$0.20         &$0.161\pm0.062$   &$0.145\pm0.031$   \\
G111.25$-$0.76         &$0.272\pm0.031$   &$0.264\pm0.022$   \\
G013.71$-$0.08         &$0.262\pm0.015$   &$0.264\pm0.018$   \\
\enddata
\tablecomments{\footnotesize
Comparison of estimated parallaxes for two methods for three representative 
examples.  Column 2 gives standard fitting results, where multiple background 
quasars are assumed to give the same parallax and effectively averaged together.  
Column 3 uses ``Method-2'' (described in the text), where the position of an 
artifical quasar at the target maser position is generated at each epoch and 
used for fitting parallax; these uncertaintes have been inflated by 10\% as 
discussed in Section \ref{sect:considerations} .
               }
\label{table:parallaxes}
\end{deluxetable}

Method-2 assumes that the $x$-components of the measured position differences
depend only on the $x$-separations of the quasars from the maser on the sky 
(and the same for the $y$-components):   
$$\Qxij = S^x_x\Offsetx + C^x~~,\eqno(3)$$
and
$$\Qyij = S^y_y\Offsety + C^y~~.\eqno(4)$$
This method has only two parameters per coordinate and hence only requires 
two quasars at a minimum to work, and if more quasars are available one will
get a more robust artificial quasar data.  This method also allows for linear 
distributions of quasar on the sky, which in principle could yield excellent 
artificial quasar data.  Using this method, we generated artificial quasar data 
at each epoch and used these to fit for parallax and proper motion of the masers.   
The parallax values generated in this fashion are indicated in the left panels of
Fig.~\ref{fig:parallax_on_sky}. Based on a visual inspection of the three cases 
presented, these give reasonable values.  

Table \ref{table:parallaxes} lists the parallax results for the three examples 
shown in Fig. \ref{fig:parallax_on_sky} both for a standard fitting approach, 
where all background quasars are assumed to yield the same parallax for the target 
maser, and for the fitting approach of Method-2 described above.  In both cases, 
the parallax uncertainties are estimated from the magnitude of the post-fit residuals.  
For these three examples, Method-1 and Method-2 returned
parallax values within a few micro-arcseconds of each other and nearly 
identical formal uncertainties.  However, in some other cases, Method-1 gave 
larger formal uncertainties.  Because of this, and because Method-1 requires 
solving for more parameters, we adopted the simpler Method-2 over Method-1 
for our application.

As is evident in Fig. \ref{fig:parallax_on_sky}, G030.78$+$0.20 displays an unusually 
large parallax gradient on the sky, and the parallax uncertainty from Method-2 
artificial quasar data is 50\% that of the standard method.  The source 
G111.25$-$0.76 displays a moderate parallax gradient on the sky, and the 
artifical quasar uncertainty is 70\% that of the standard estimate.  
In the final example, G013.71$-$0.08, little or no parallax gradient is detected 
and the standard and artificial quasar methods produce similar results.   Since,
in the presence of significant parallax gradients, the artifical quasar (Method-2) 
almost always gave smaller post-fit residuals than from a simple combined fit 
using all quasars, which does not take into account possible position gradients 
on the sky, we adopted it as the method of choice for the BeSSeL Survey 6.7-GHz data.

\section{Other Considerations} \label{sect:considerations}

When we planned BeSSeL Survey observations for VLBA programs BR149 and BR198,
we scheduled the minimum number of epochs for parallax and proper motion 
observations, which would yield uncorrelated parameter estimates and the 
lowest parallax uncertainty.  This could be accomplished with four epochs 
scheduled to sample the peaks of the parallax signature in Right Ascension,
for example, for sources at low Galactic longitude we would schedule one 
observation in March, two in September, and one in the following March.   
We optimized based on the Right Ascension component of the parallax signature, 
since 1) that component has a larger amplitude than the Declination component 
for most Galactic sources, and 2) VLBA measurements are generally more accurate 
for that component than for the Declination component \citep{Reid:09a}.  

By observing three or four quasars for each maser target, we planned to have 
sufficient degrees of freedom to estimate true astrometric accuracy from 
the magnitudes of the residuals to the parallax fits.   For example, with four 
background quasars for a maser target and four epochs of observations, there
are 16 relative position measurements in the critical Right Ascension direction.
Since all background quasars should yield the same parallax and motion for
the maser, each additional quasar only adds one extra parameter, a correction 
to its position offset, while adding four extra data points.  Thus, for this 
example, there would be 16 data points and 6 parameters to solve for, yielding 
10 degrees of freedom.  

\begin{figure}[ht]
\epsscale{0.8} 
\plotone{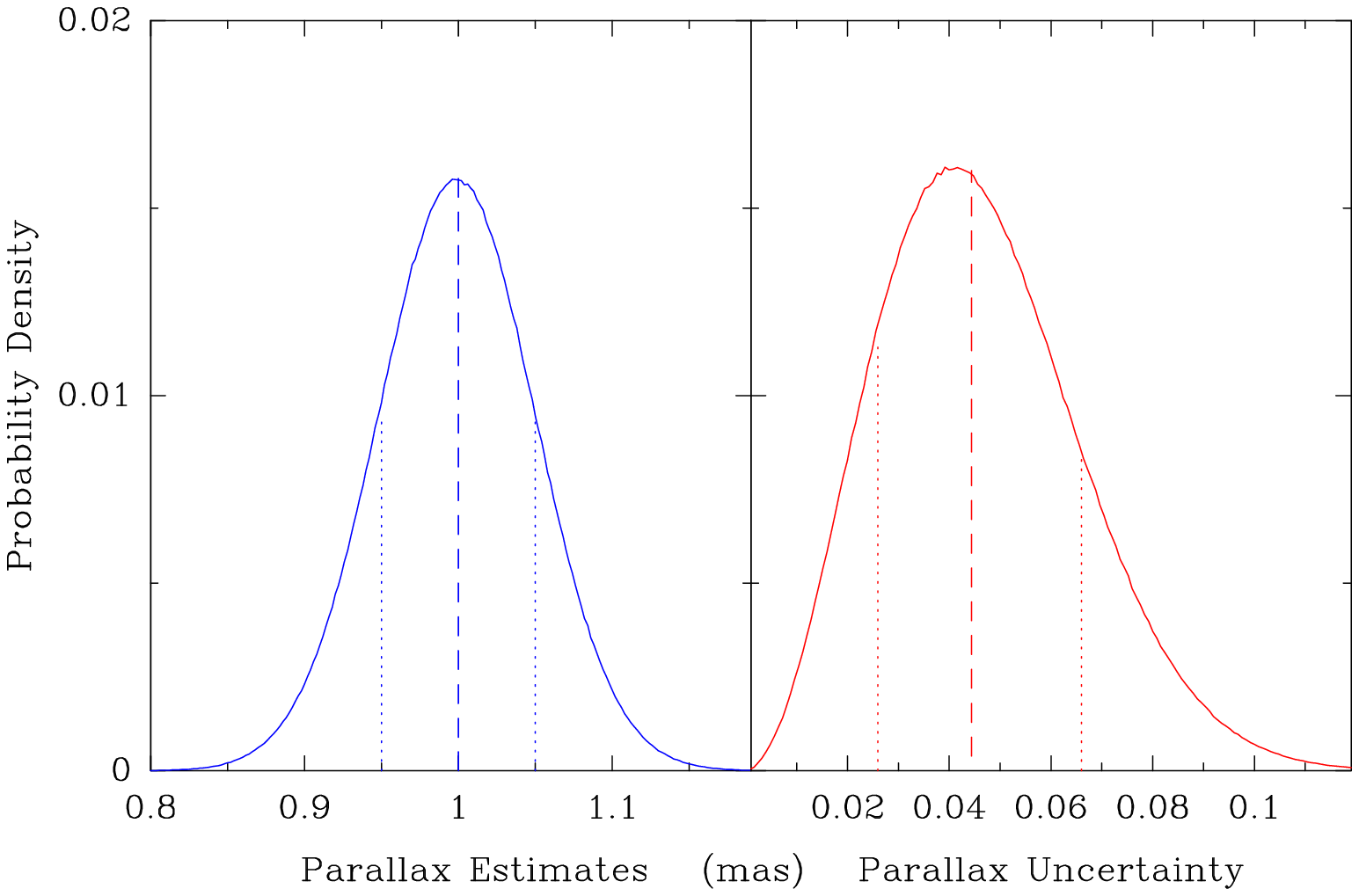}
\caption{\footnotesize
Results of $10^7$ simulations of parallax estimates which used four optimally 
sampled measurements of the Right Ascension difference between a maser and 
a background quasar.   Data were
simulated for a parallax of 1.00 mas and single-epoch position measurement 
uncertainty of 0.10 mas, which should yield a fitted parallax uncertainty 
of 0.05 mas.   Binned histograms tracing probability density functions (PDF) are
shown with solid lines; median values are indicated with dashed lines and the 
68\% confidence ranges are indicated with dotted lines.
{\it Left Panel:} PDF for the parallax trials, which return the expected
estimates of parallax.
{\it Right Panel:} PDF for the formal parallax uncertainties, scaled by
the post-fit residuals to give unity reduced chi-squared values.
This PDF is asymmetric with mean and median values for the uncertainty 
$\approx10$\% below the correct value of 0.05 mas and a tail toward
larger values.  The symmetric 68\% confidence range is 0.026 to 0.066 mas
(uncorrected for bias).
        }
\label{fig:parallax_uncertainties}
\end{figure}

However, in order to mitigate the effects of position gradients across
our source groupings, we used up most of our degrees of freedom to generate 
the artificial quasar data.  Fits to this data for four epochs have only one 
degree of freedom when the Right Ascension data dominates.  
This in no way biases the parallax estimates, but it leads to
uncertain estimates for the {\it uncertainty} in the parallax.  
Fig.~\ref{fig:parallax_uncertainties} shows the results of $10^7$ simulated
parallax fits with only one degree of freedom (\ie using 4 optimally sampled 
data points and 3 parameters).  The true parallax value was set at 1.00 mas and 
the Right Ascension ($1\sigma$) position uncertainty was set to 0.10 mas.  
These values should lead to a true parallax uncertainty of 0.05 mas, since 
we actually measure twice the parallax amplitude.  The left panel of the figure
shows that we retrieve the correct values and there is no bias in the 
parallax estimate.   The right panel shows the distribution of formal parallax 
uncertainties, estimated from the scatter in the post-fit residuals.  The 
probability density is asymmetric with mean and median values of $\approx0.0444$ 
mas (about 10\% below the correct value of 0.05) and a tail to larger values.
Being conservative, we have inflated our parallax uncertainties accordingly 
to remove this slight bias.  Following this correction, the symmetric 68\% 
confidence range for the parallax uncertainties for this example spans from 
0.029 to 0.073 mas about the correct value of 0.05 mas.  Thus, the 
{\it uncertainty in the uncertainty} can be expected to be significant 
($\pm44$\%).  In conclusion, even though the parallax estimate is unbiased, 
one should exercise caution when using the estimate of its uncertainty from 
our 6.7-GHz parallax measurements.

\vskip 0.5truein\noindent 
{\it Facilities:}  \facility{VLBA}

\end{document}